\documentclass[review]{elsarticle}

\usepackage{lineno,hyperref,amsmath}
\usepackage{graphicx,subcaption,epstopdf}
\usepackage{color}

\usepackage[doublespacing]{setspace}

\usepackage{booktabs} 
\modulolinenumbers[1]

\begin{document}
\begin{frontmatter}

\title{Fast chaos indicator from auto-differentiation for dynamic aperture optimization}

\author[a]{J. Qiang\corref{author}}
\ead{jqiang@lbl.gov}
\author[b]{J. Wan}
\author[c]{A. Qiang}
\author[b]{Y. Hao}

\cortext[author] {Corresponding author.}

\address[a]{Lawrence Berkeley National Laboratory, Berkeley, CA 94720, USA}
\address[b]{Facility for Rare Isotope Beams, Michigan State University, East Lansing, 48824, USA}
\address[c]{Stanford University, 450 Jane Stanford Way,
Stanford, CA 94305, USA}

\begin{abstract}
  Automatic differentiation provides an efficient means of computing derivatives of complex functions with machine precision, thereby enabling differentiable simulation. In this work, we propose the use of the norm of the tangent map, obtained from differentiable tracking of particle trajectories, as a computationally efficient indicator of chaotic behavior in phase space. In many cases, a one‑turn or few‑turn tangent map is sufficient for this purpose, significantly reducing the computational cost associated with dynamic aperture optimization. As an illustrative application, the proposed indicator is employed in the dynamic aperture optimization of an ALS‑U lattice design.
\end{abstract}

\end{frontmatter}


\section{INTRODUCTION} 

The dynamic aperture (DA) is a critical parameter in the design of circular particle accelerators, as it defines the region within which a charged particle remains stable over a specified number of turns. Particles whose initial positions extend beyond the DA boundary are lost from the system.

A conventional approach to determining the DA involves tracking a large number of macroparticles with different initial positions over many turns~\cite{borland,sun11,yang,gao,huang}. The initial positions of the outermost stable macroparticles after tracking are then used to delineate the DA boundary. While accurate, this brute‑force method is computationally intensive and time‑consuming, making it a challenge for efficient DA optimization.

In recent years, machine learning–based approaches have been proposed to accelerate DA optimization~\cite{li2018,kran,wan2022,croce}. These methods construct a surrogate model from a relatively small set of training data, which is then employed to guide the search for configurations yielding a larger DA during the optimization process. The surrogate model is iteratively updated as optimization progresses, thereby improving its predictive accuracy. Although effective in reducing computation time, the performance of these methods depends strongly on the quality and representativeness of the initial training data, and they may require retraining for lattices exhibiting substantially different dynamical behavior.

Other efficient methods, such as frequency map analysis (FMA) and the reversibility error method (REM), have also been studied in the accelerator community to accelerate DA estimation\cite{laskar1,laskar2,steier,sun,yonnis,xu,hwang,li2}. In the FMA approach, changes in the particle trajectory frequency, quantified as the tune diffusion rate, {are employed as indicators of chaotic motion in the presence of high-order multipole nonlinear fields}. A large tune diffusion rate corresponds to strongly chaotic behavior;
{in such cases, particles within chaotic regions are typically lost quickly in a lattice with misalignment field errors}.
In the REM approach, a particle is tracked forward for a specified number of turns and then tracked backward to its initial time or longitudinal coordinate. The separation between the initial and final phase‑space coordinates provides a measure of sensitivity to perturbations from numerical round‑off errors. For regular trajectories, this separation grows as a power law with respect to the number of steps, whereas for chaotic trajectories it grows exponentially.

Automatic differentiation (AD) is a computational technique that enables the efficient evaluation of derivatives of complex functions with respect to a given set of parameters, without resorting to numerical approximation or symbolic differentiation. In the artificial intelligence and machine learning (AI/ML) community, AD has been widely employed for training the parameters of neural networks\cite{ad,pytorch,tensorflow}. More recently, AD has found applications in particle accelerator physics, where it has been utilized to quantify the sensitivity of simulation results to accelerator lattice parameters and to expedite optimization procedures\cite{roussel2022,roussel2023b,qiang2023,cheetah,wan2}.

In particle dynamical systems, a hallmark of chaotic trajectories is the pronounced sensitivity of the final particle position to small variations in the initial coordinates. This sensitivity can be quantified by computing the derivatives of the final positions with respect to the initial coordinates, which form the Jacobian matrix (also referred to as the tangent map). Using AD, these derivatives and the corresponding tangent map can be obtained directly during simulation with machine precision. In this work, we propose employing the norm of the tangent map, computed via differentiable particle tracking, as a fast and reliable indicator of chaotic trajectories. By exploiting one‑turn or few‑turn tangent matrices, this method significantly reduces computation time while maintaining accuracy, thereby enabling more efficient DA optimization.

The remainder of this paper is organized as follows. Section~2 introduces the forward mode of automatic differentiation. Section~3 describes the tangent map norm indicator. Section~4 presents a benchmark example based on the Hénon–Heiles potential. Section~5 demonstrates the application of the proposed fast indicator to dynamic aperture optimization. Finally, Section~6 summarizes the conclusions.

\section{Forward mode automatic differentiation}

The automatic differentiation (AD) module used in this study is based on the Truncated Power Series Algebra (TPSA) method~\cite{berz} but is limited to first derivatives~\cite{qiangipac25}. 
This restriction substantially reduces the 
 computational complexity of the arithmetic operations
 and decreases execution time by orders of magnitude, while also simplifying the module’s implementation. 
In the artificial intelligence and machine learning (AI/ML) community, this approach is equivalent to the \emph{dual‑number} implementation of forward‑mode AD~\cite{adwiki}.

In TPSA, the computation of a function’s derivatives with respect to its variables is reformulated as the evaluation of the function on a vector‑like TPSA variable according to predefined algebraic rules. For a function $f(x_1,x_2,\ldots,x_n)$ with $n$ independent variables and its first derivatives, we define a TPSA vector variable
\[
F = \left(f, f_{x_1}, f_{x_2}, \ldots, f_{x_n}\right),
\]
where $f$ is the function value and $f_{x_i} \equiv \partial f / \partial x_i$ is the partial derivative with respect to variable $x_i$. For the independent variable $x_i$, the corresponding TPSA vector is
\[
X_i = \left( x_i, 0, 0, \ldots, 1, \ldots \right),
\]
where the value $1$ appears at the $(i+1)^{\mathrm{th}}$ element, indicating the derivative with respect to $x_i$.

The elementary operations for two TPSA vectors $F$ and $G$ are defined as follows:

\paragraph{Addition:}
\begin{equation}
F + G = \left(f+g, \; f_{x_1}+g_{x_1}, \; f_{x_2}+g_{x_2}, \ldots, f_{x_n}+g_{x_n}\right),
\end{equation}

\paragraph{Multiplication:}
\begin{equation}
FG = \left(fg, \; g f_{x_1} + f g_{x_1}, \; g f_{x_2} + f g_{x_2}, \ldots, g f_{x_n} + f g_{x_n}\right),
\end{equation}

\paragraph{Division:}
\begin{equation}
\frac{F}{G} =
\left(\frac{f}{g},\;
\frac{f_{x_1}g - g_{x_1}f}{g^{2}},\;
\frac{f_{x_2}g - g_{x_2}f}{g^{2}},\;
\ldots,\;
\frac{f_{x_n}g - g_{x_n}f}{g^{2}}\right),
\end{equation}

\paragraph{Function mapping:}
\begin{equation}
h(F) = \left(h(f), \; h_f f_{x_1}, \; h_f f_{x_2}, \ldots, h_f f_{x_n}\right),
\end{equation}
where $h_f \equiv \partial h / \partial f$.

By implementing these operational rules, we define a specialized TPSA vector data type along with corresponding arithmetic and standard mathematical functions. In the differentiable particle‑tracking program, the particle phase‑space coordinates—normally stored as double‑precision variables—are instead declared using this data type. As a result, the Jacobian matrix, or tangent map, is obtained automatically through the simulation without requiring separate derivative calculations.

\section{Norm of tangent map as chaos indicator}

For a particle moving inside an accelerator, the long-term behavior of its trajectory can be characterized using the maximal Lyapunov exponent\cite{liapounoff}. For regular motion, the separation between two initially nearby particles grows linearly with respect to the distance $s$ when averaged over long distances. For chaotic motion, the separation grows exponentially with $s$, i.e.,
\begin{equation}
    ||\Delta \zeta(s)|| \propto ||\Delta \zeta(0)|| \, \exp(\lambda s),
\end{equation}
where $||\Delta \zeta(0)||$ denotes the initial separation, $||\Delta \zeta(s)||$ the separation after traveling a distance $s$, and $\lambda$ the maximal Lyapunov exponent. This exponent is formally defined as
\begin{equation}
    \lambda = \lim_{s \to \infty} \, \lim_{\Delta \zeta(0) \to 0} 
    \frac{1}{s} \ln \left( \frac{||\Delta \zeta(s)||}{||\Delta \zeta(0)||} \right).
\end{equation}

The evolution of the deviation vector $\Delta \zeta(s)$ can be expressed in terms of the tangent map (Jacobian matrix) $M(s)$ as
\begin{equation}
    \Delta \zeta(s) = M(s) \, \Delta \zeta(0).
    \label{eq7}
\end{equation}
Thus, the relative separation becomes
\begin{equation}
    \frac{||\Delta \zeta(s)||}{||\Delta \zeta(0)||} 
    = \frac{||M(s) \Delta \zeta(0)||}{||\Delta \zeta(0)||}.
\end{equation}

Here, $M(s)$ is the tangent map computed at distance $s$ with respect to the
initial deviation. Such a tangent map can be obtained automatically using the forward-mode AD tracking procedure described in Sec.~2. By noting that
\begin{equation}
    ||M(s)\,\Delta \zeta(0)|| \leq ||M(s)|| \, ||\Delta \zeta(0)||,
\end{equation}
we obtain the bound
\begin{equation}
    \frac{||\Delta \zeta(s)||}{||\Delta \zeta(0)||} \leq ||M(s)||.
\end{equation}

For a chaotic trajectory, where the separation between initially nearby trajectories grows exponentially with $s$, the norm $||M(s)||$ of the tangent map will likewise grow exponentially. 
Furthermore, if the norm $||M(s)||$ grow exponentially, at least one element of the matrix $M(s)$ should
grow exponentially. From Eq.~\ref{eq7}, the separation $||\Delta \zeta(s)||$ will grow exponentially.
These observations suggest that the norm of the tangent map can serve as a computationally efficient indicator of chaos in particle motion.

\section{Test of the Indicator Using a H\'enon--Heiles Potential}

We first tested the proposed fast chaos indicator, based on the norm of the tangent map, using the four-dimensional H\'enon--Heiles problem studied previously~\cite{henon}. The Hamiltonian for this system is given by
\begin{equation}
    H(x,p_x,y,p_y) = \frac{p_x^2 + p_y^2}{2} + \frac{x^2 + y^2}{2} + x^2 y - \frac{y^3}{3}.
\end{equation}
Let $\zeta = (x, p_x, y, p_y)$ denote the four-dimensional phase-space coordinate vector. Hamilton’s equations can then be expressed as
\begin{equation}
    \frac{d\zeta}{dt} = -[H, \zeta],
\end{equation}
where $[\cdot, \cdot]$ denotes the Poisson bracket.  

A formal solution to the above equation after a single time step $\tau$ is
\begin{equation}
    \zeta(\tau) = \exp\left(-\tau \, (:H:)\right) \zeta(0),
\end{equation}
where the Lie operator $:H:$ is defined by $:H: g = [H, g]$ for an arbitrary function $g$~\cite{dragt}.  

For a Hamiltonian separable into two parts,
\[
    H = H_1 + H_2, \quad H_1 = \frac{p_x^2 + p_y^2}{2} + \frac{x^2 + y^2}{2}, \quad 
    H_2 = x^2 y - \frac{y^3}{3},
\]
an approximate solution can be obtained using a second-order symmetric splitting formula:
\begin{eqnarray}
    \zeta(\tau) & = & \exp\left(-\frac{\tau}{2} :H_1:\right) \exp\left(-\tau :H_2:\right) 
    \exp\left(-\frac{\tau}{2} :H_1:\right) \zeta(0) \nonumber \\
    & & + \ \mathcal{O}(\tau^3).
    \label{ham_split}
\end{eqnarray}
Defining $\exp\left(-\frac{\tau}{2} :H_1:\right) \equiv \mathcal{M}_1(\tau/2)$ and  
$\exp\left(-\tau :H_2:\right) \equiv \mathcal{M}_2(\tau)$, the single-step map becomes
\begin{eqnarray}
    \zeta(\tau) & = & \mathcal{M}(\tau) \zeta(0) \nonumber \\
    & = & \mathcal{M}_1(\tau/2) \, \mathcal{M}_2(\tau) \, \mathcal{M}_1(\tau/2) \zeta(0) + \mathcal{O}(\tau^3).
    \label{map}
\end{eqnarray}

In the differentiable simulation, the vector $\zeta$ is declared as a differentiable variable $D\zeta$ using the TPSA-based automatic differentiation data type described in Sec.~2. The corresponding update rule is then
\begin{eqnarray}
    D\zeta(\tau) & = & \mathcal{M}_1(\tau/2) \, \mathcal{M}_2(\tau) \, \mathcal{M}_1(\tau/2) D\zeta(0).
    \label{damap}
\end{eqnarray}
After integrating over a specified time interval $T$, the simulation yields both the phase-space coordinates $\zeta(T)$ and the Jacobian (tangent map) $M(T)$.  

In this test, three matrix norms of $M(T)$ were evaluated as potential chaos indicators:  
(i) the $p=1$ norm (maximum absolute column sum),  
(ii) the $p=\infty$ norm (maximum absolute row sum), and  
(iii) the Frobenius norm (square root of the sum of absolute squares of all matrix elements).

\begin{figure}[!htb]
\centering
\includegraphics[width=8.0cm]{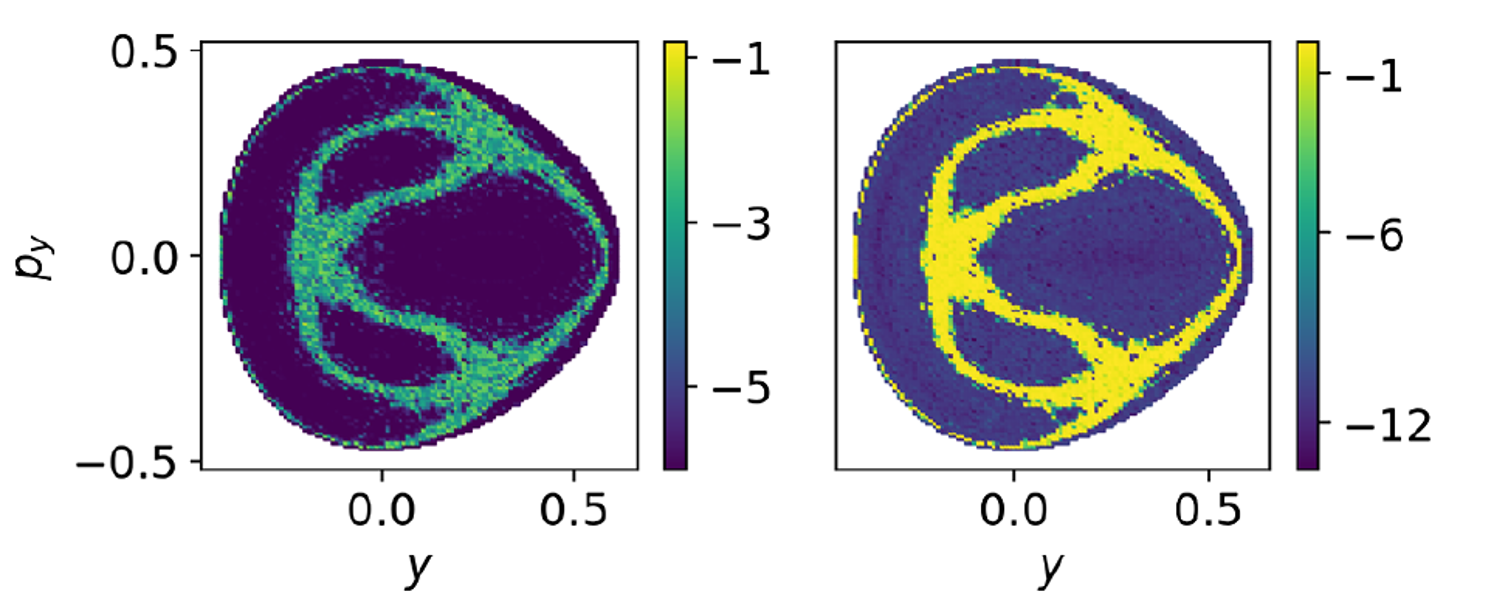} \\[2mm]
\includegraphics[width=3.5cm]{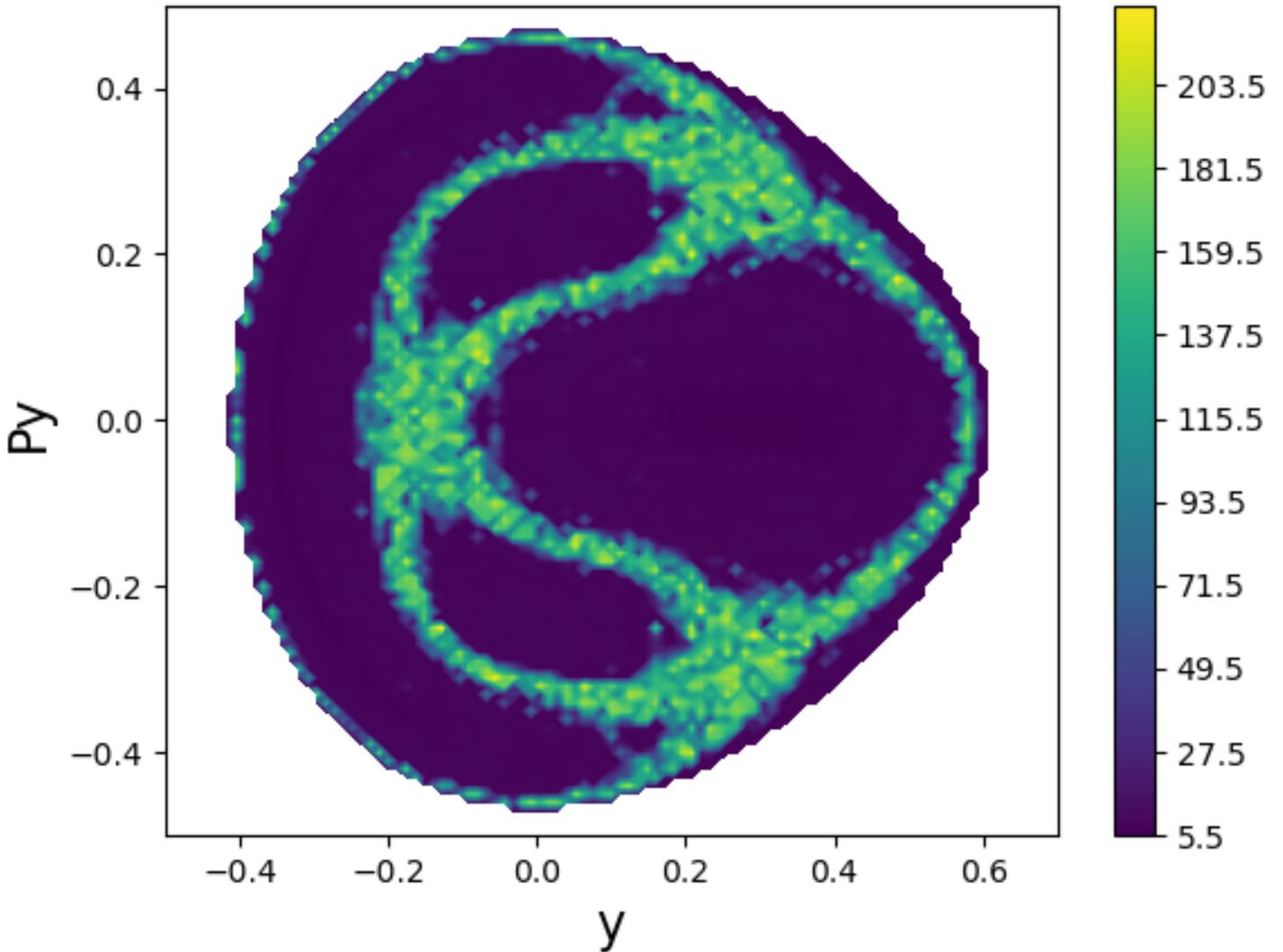} \hspace{1mm}
\includegraphics[width=3.5cm]{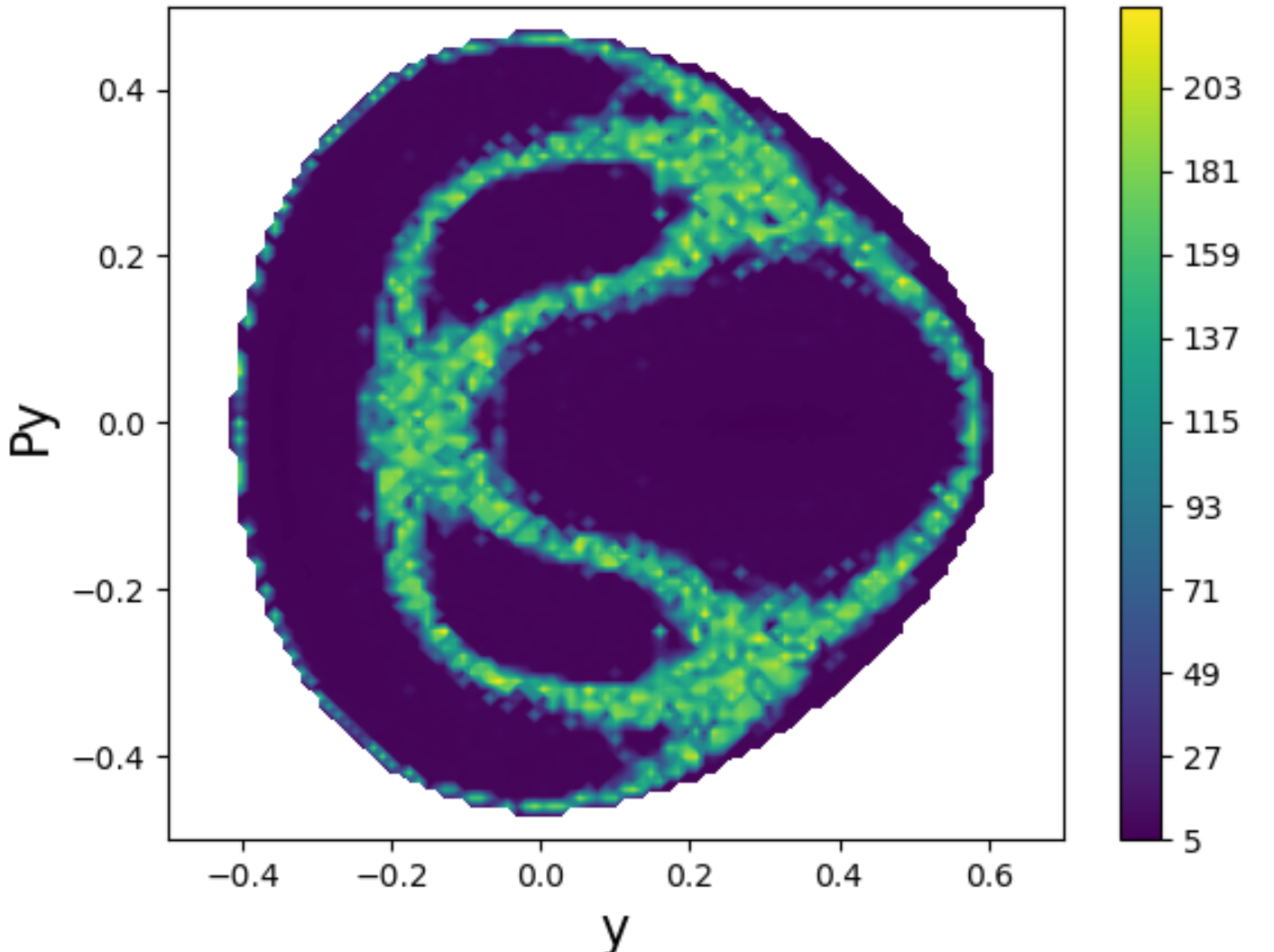} \hspace{1mm}
\includegraphics[width=3.5cm]{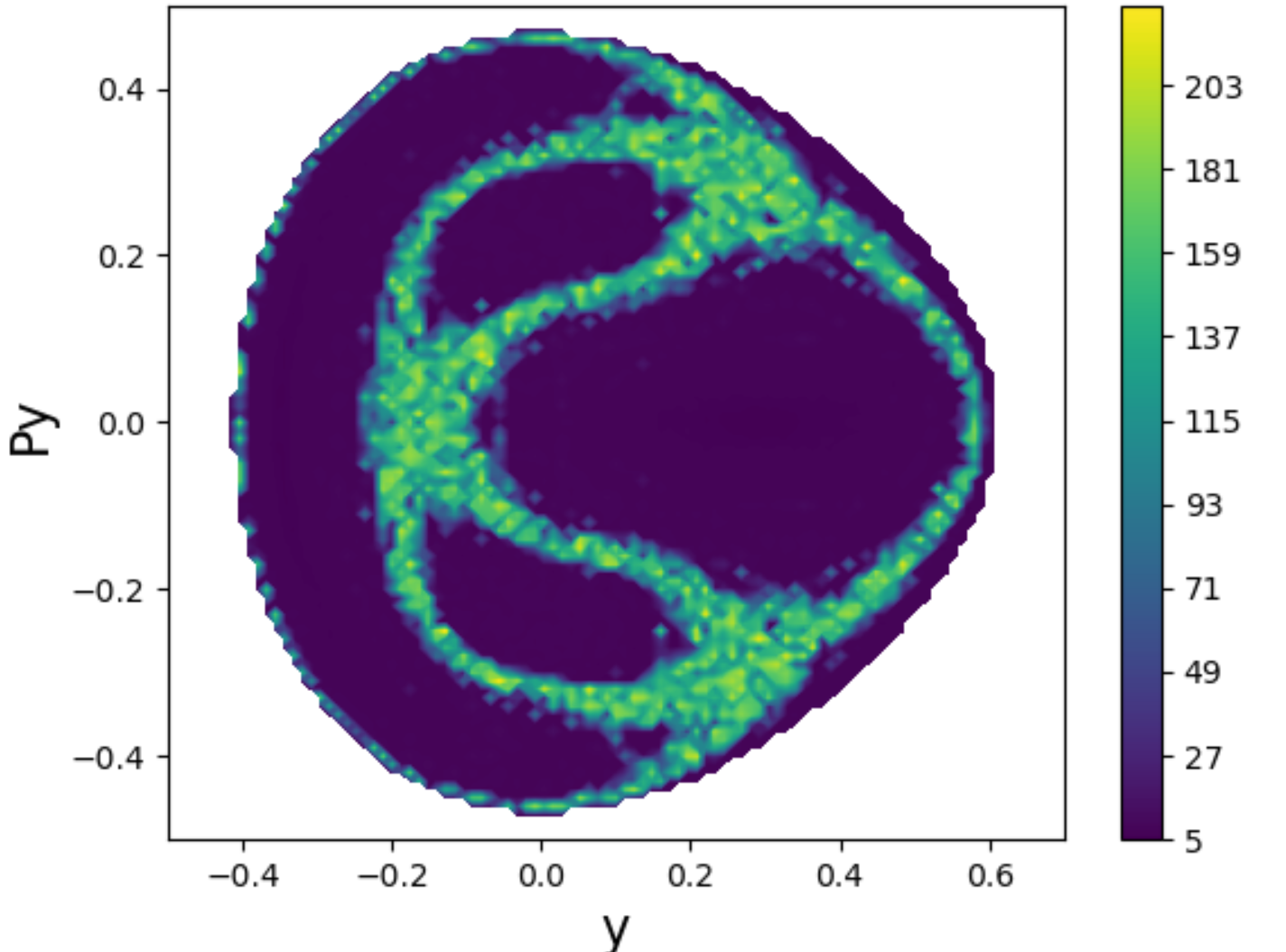}
\caption{
Dynamic aperture in the $y$--$p_y$ plane obtained using FMA (top left) and REM (top right) indicators~\cite{hwang}, and using the $p=1$ (bottom left), $p=\infty$ (bottom middle), and Frobenius (bottom right) norms of the tangent map computed via differentiable tracking.}
\label{fig1}
\end{figure}

For benchmarking, we adopted the same energy level $H = 1/9$ and the condition $x=0$ as used in Ref.~\cite{hwang}. Figure~\ref{fig1} shows the dynamic aperture in the $y$--$p_y$ plane after a time interval $T=4096$, using each of the three norm indicators. For comparison, results obtained using the FMA and REM indicators~\cite{hwang} for the same problem are also shown.  

The phase-space structures produced by the norm-based indicators agree closely with those obtained using FMA and REM. All three norm indicators yield nearly identical phase-space structures and dynamic aperture boundaries. In the following section, we will employ the Frobenius norm as the preferred indicator for dynamic aperture optimization.

\section{Application to Dynamic Aperture Optimization}

As an illustration of the use of the tangent map norm from automatic differentiation as a fast chaos indicator, we apply this method to a synchrotron light source design for the ALS upgrade (ALS-U)~\cite{alsu}. The ALS-U is an upgrade of the existing third-generation synchrotron light source, the Advanced Light Source (ALS), to a fourth-generation diffraction-limited light source employing a multi-bend lattice. The upgrade aims to significantly reduce the horizontal emittance of the electron beam and enhance brightness by orders of magnitude.

\begin{figure}[!htb]
\centering
\includegraphics[width=12.0cm]{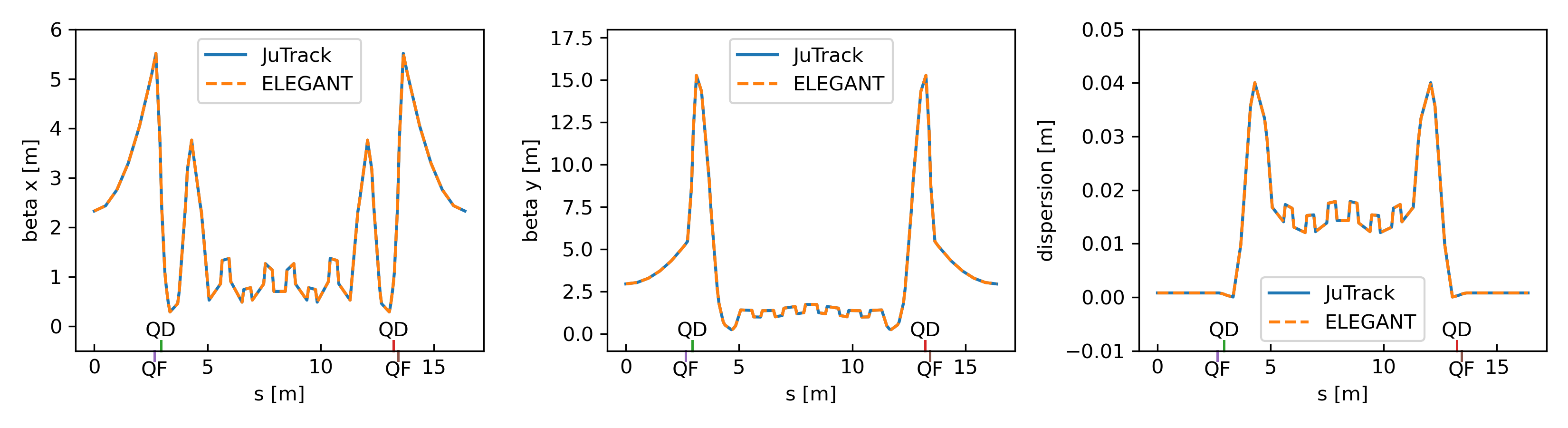}
\caption{
Evolution of the horizontal beta function (left), vertical beta function (middle), and dispersion (right) over one of the $12$ cells in the ALS-U ring, as computed with \texttt{JuTrack} and \texttt{Elegant}. The positions of the two quadrupoles---QF and QD---used to control the machine working tunes are indicated.
}
\label{fig2}
\end{figure}

\begin{figure}[!htb]
\centering
\includegraphics[width=5.0cm]{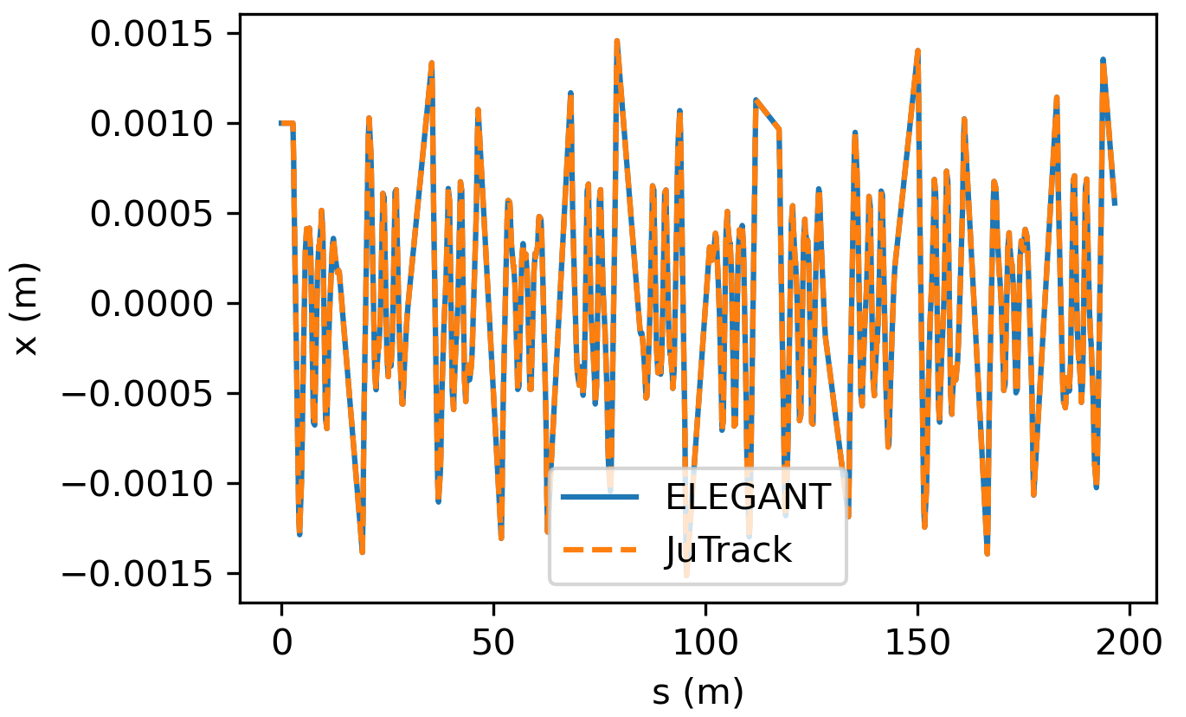}
\includegraphics[width=5.0cm]{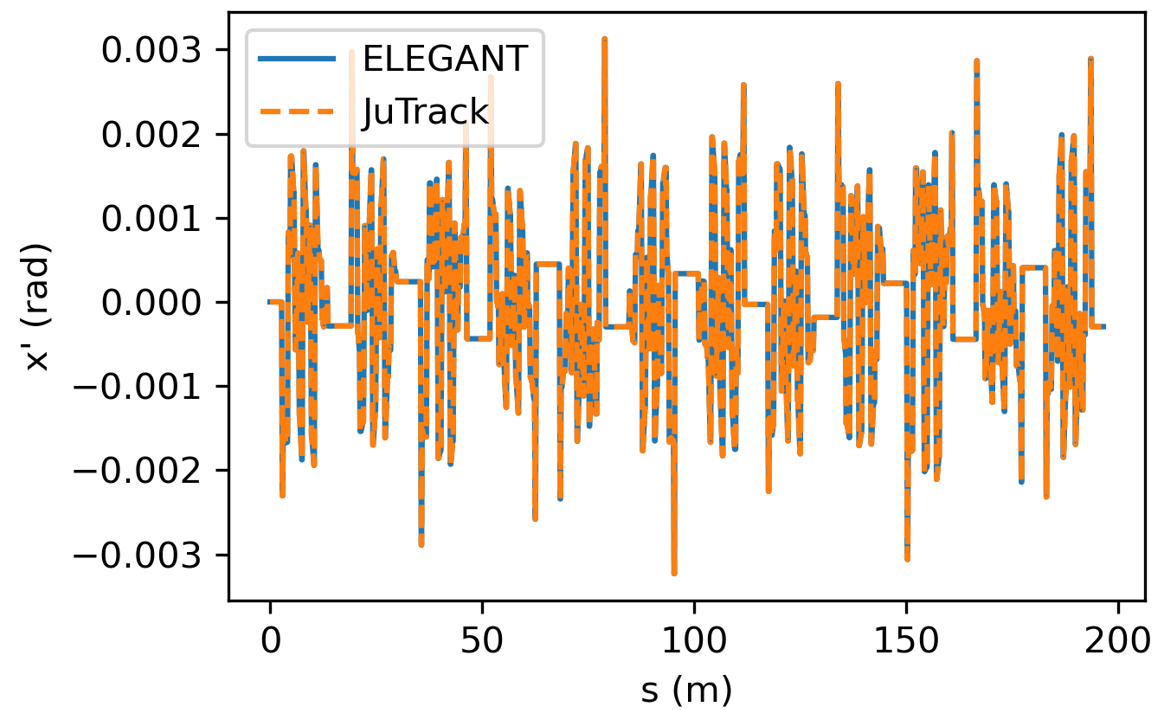}
\includegraphics[width=5.0cm]{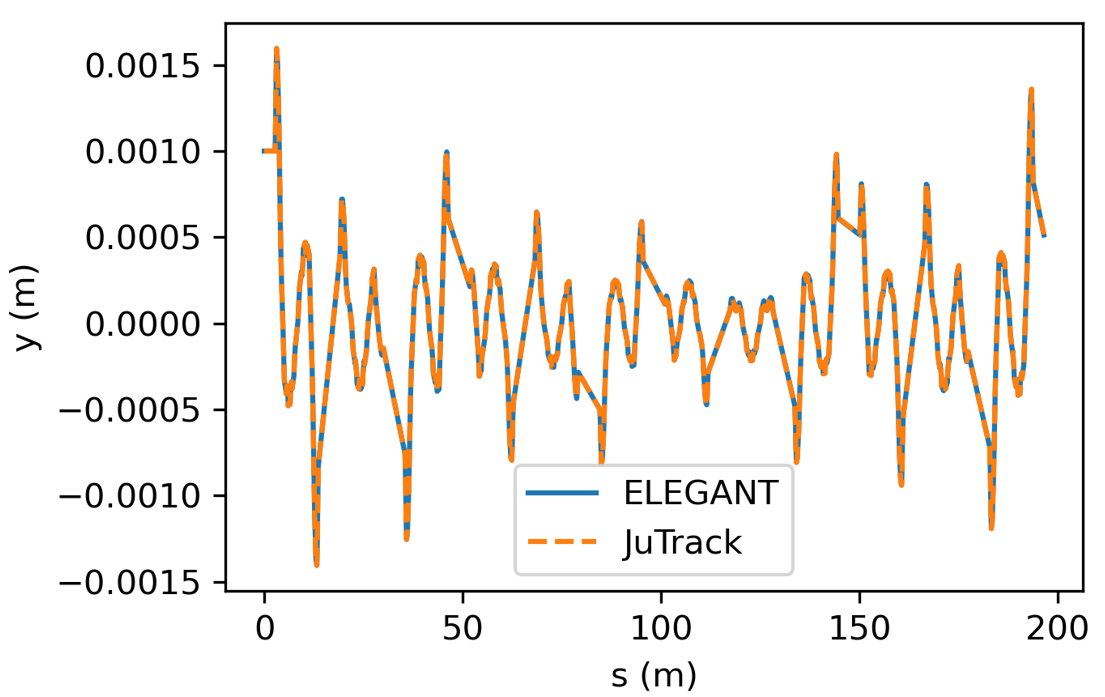}
\includegraphics[width=5.0cm]{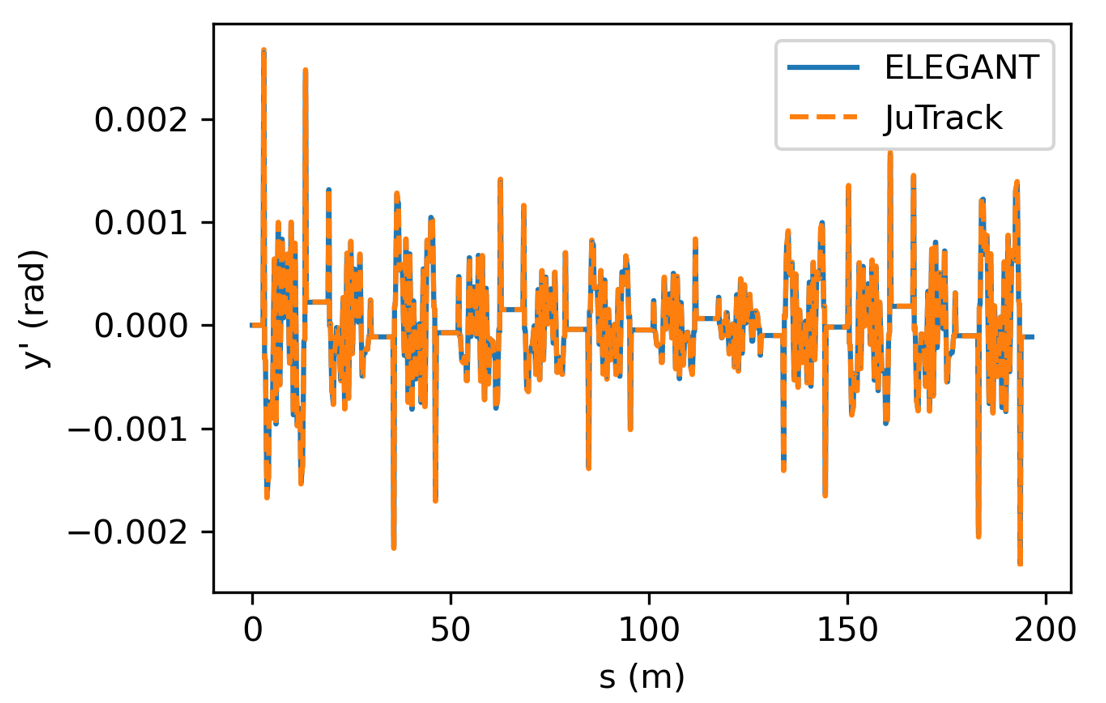}
\caption{
Element by element evolution of the $X$ (top left), $X'$ (top right), $Y$ (bottom left) and $Y'$ (bottom right) as computed with \texttt{JuTrack} and \texttt{Elegant} for a particle starting
at coordinates $(0.001,0,0.001,0,0,0)$.
}
\label{fig2track}
\end{figure}

In this study, we use one version of the ALS-U lattice design for demonstration purposes; it does not represent the latest official ALS-U configuration. Particle tracking is performed using the differentiable tracking code \texttt{JuTrack}~\cite{wan2} to determine the dynamic aperture. JuTrack is a modern, AD-enabled accelerator tracking software package written in the Julia programming language. It has been thoroughly benchmarked against widely used accelerator codes, such as MADX, demonstrating consistent accuracy in particle tracking and optics functions. Furthermore, JuTrack provides accurate derivatives with respect to any pre-defined simulation parameters through automatic differentiation. These AD results have been fully validated against finite difference methods. Figure~\ref{fig2} compares the horizontal, vertical beta functions, and horizontal dispersion evolution through one of $12$ cells of the nominal ALS-U lattice obtained {using the first-order TPSA method in the \texttt{JuTrack} and the \texttt{Elegant}~\cite{elegant} simulation code}. The results show excellent agreement between the two codes. 
The nominal working tunes for the design are $41.374$ (horizontal) and $20.379$ (vertical).
In addition to the optics function benchmarks, we compared single-particle tracking results from both codes. Figure~\ref{fig2track} shows the element-by-element evolution of the transverse phase space through the accelerator using JuTrack and Elegant. As demonstrated, both codes show excellent agreement.


Since the three norms of the tangent map tested in Sec.~4 yield nearly identical phase-space structures, we use the Frobenius norm of the tangent map after one turn as a fast indicator for ALS-U dynamic aperture optimization.  

\begin{figure}[!htb]
\centering
\includegraphics[width=4.8cm]{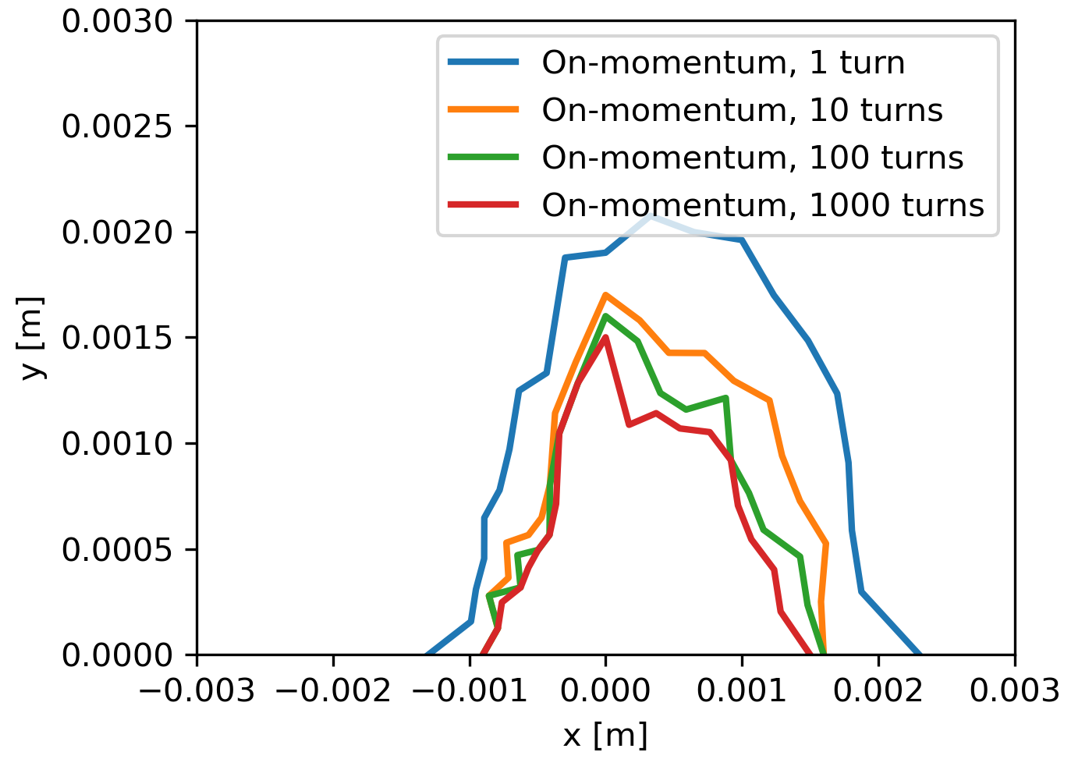}
\includegraphics[width=5.0cm]{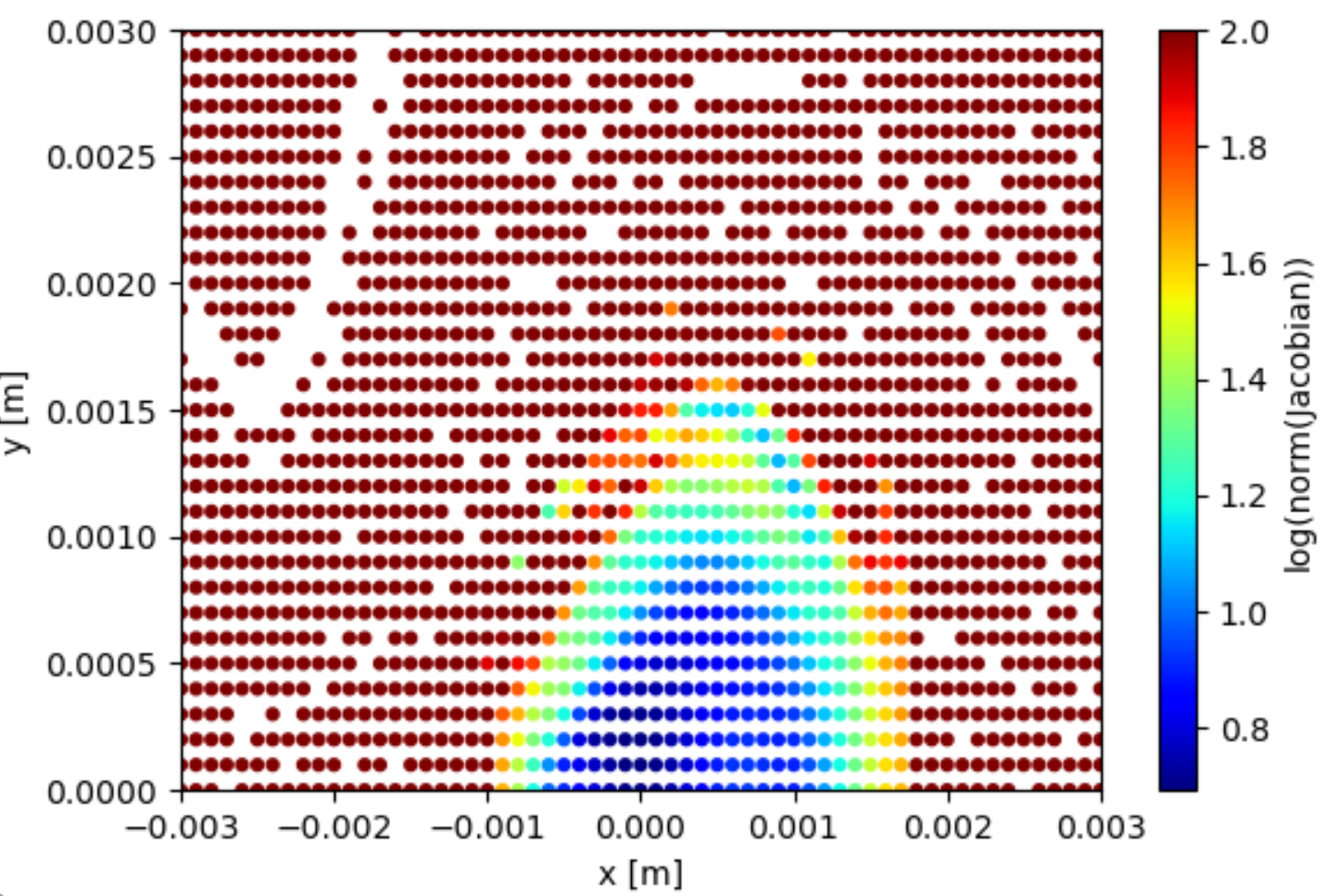}
\caption{
Left: boundary of surviving particles in the $x$--$y$ plane after 1, 10, 100, and 1000 turns of tracking.  
Right: spatial distribution of the log Frobenius norm of the tangent map after one-turn differentiable tracking with zero momentum deviation in the nominal ALS-U lattice.}
\label{fig3}
\end{figure}

Figure~\ref{fig3} shows the boundary lines in the $x$--$y$ plane for particles that survive after tracking through $1$, $10$, $100$, and $1000$ turns of the ALS-U lattice with zero momentum deviation, along with the spatial distribution of the logarithm of the Frobenius norm of the tangent map after one-turn differentiable tracking. Here, the tangent map is based on the normalized coordinates. The direct particle tracking boundary converges at $1000$ turns, and the spatial pattern of the log Frobenius norm after one turn closely matches the $1000$-turn result. This suggests that the one-turn log Frobenius norm can serve as a fast, reliable indicator for dynamic aperture optimization.

\begin{figure}[!htb]
\centering
\includegraphics[width=5.0cm]{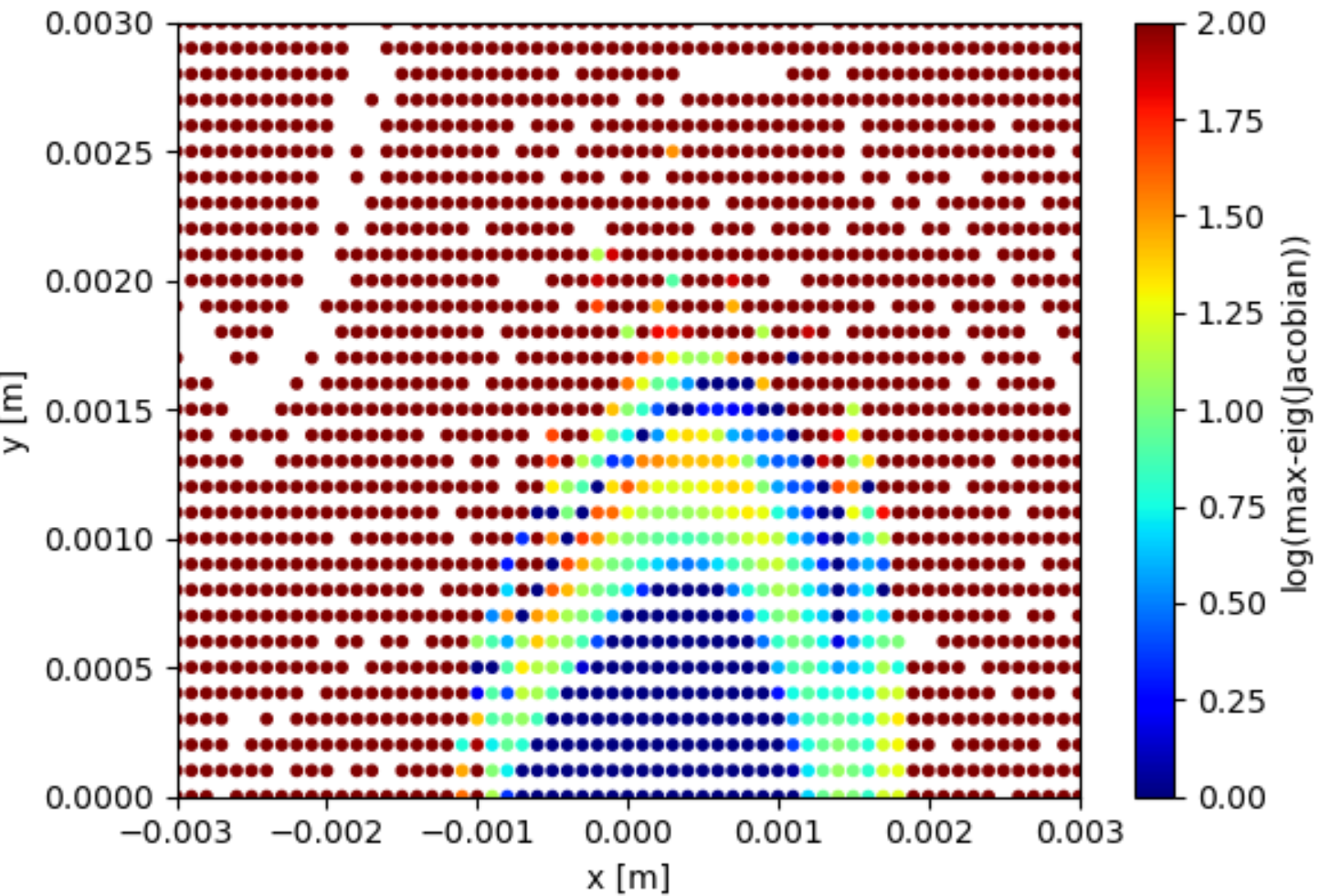}
\caption{
Spatial distribution of the log of the maximum eigenvalue of the one-turn tangent map from differentiable tracking with zero momentum deviation in the nominal ALS-U lattice.}
\label{fig4}
\end{figure}

In the limit of infinite distance (or time), the logarithm of the maximum eigenvalue of the tangent map yields the Lyapunov exponent and can also act as an indicator of chaotic trajectories once the tangent map is known. Figure~\ref{fig4} shows the distribution of the log of the maximum eigenvalue of the one-turn tangent map, which resembles the Frobenius norm distribution. However, computing eigenvalues is more expensive than computing the Frobenius norm, and we therefore adopt the latter for optimization.

In this optimization example, the dynamic aperture area is maximized with respect to the two quadrupole strengths, $k_1$ and $k_2$, which control the tunes of the ALS-U accelerator, as shown in Fig.~\ref{fig2}. The objective function is defined using the one-turn logarithmic Frobenius norm in the $x$--$y$ plane. The rectangular computational domain $[-0.003,\,0.003] \times [0,\,0.003]$~m is discretized into a uniform $61 \times 31$ grid, and the one-turn log Frobenius norm is evaluated at each grid point.

Along the $x$-axis, the $y$ coordinate is scanned from zero until the log Frobenius norm exceeds a threshold value of one. The choice of a threshold value of one is based on the rapid color change shown in Figure~\ref{fig3} and is guided by the shape of the dynamic aperture from the 1,000-turn direct tracking.
We also performed an optimization using a threshold value of 1.2 instead of one. The resulting optimized parameters were not very sensitive to this change in the threshold value. The area beneath the resulting boundaries defines the estimated dynamic aperture. To encourage symmetry, areas for $x>0$ and $x<0$ are computed separately, and the absolute difference between the two is subtracted from the total area $A$ as a penalty. The objective function is therefore
\begin{equation}
    f(k_1, k_2) = A - \left| A_+ - A_- \right|,
\end{equation}
where $A_+$ and $A_-$ denote the dynamic aperture areas on the $x>0$ and $x<0$ sides, respectively. The quadrupole strengths are restricted to the range $13.0$--$14.0$~m$^{-2}$, the tunes are constrained between $0.1$ and $0.9$ to avoid significant beta beat, and
the linear chromaticities are limited below $3$.

The optimization is performed using the stochastic evolution-based parallel multi-objective optimizer \texttt{PVPmoo}~\cite{qiangpvp}, which combines an adaptive unified differential evolution algorithm~\cite{qiangde} with a real-coded genetic algorithm~\cite{deb}.  
Differential evolution is used with a $90\%$ probability for its fast convergence, and the genetic algorithm with a $10\%$ probability to improve its global search capability by random mutation.  
The initial population consists of $512$ candidates generated via quasi-random sampling in the $[13,14]$ range.  
The optimization converges after $27$ generation evolution.

\begin{figure}[!htb]
\centering
\includegraphics[width=5.0cm]{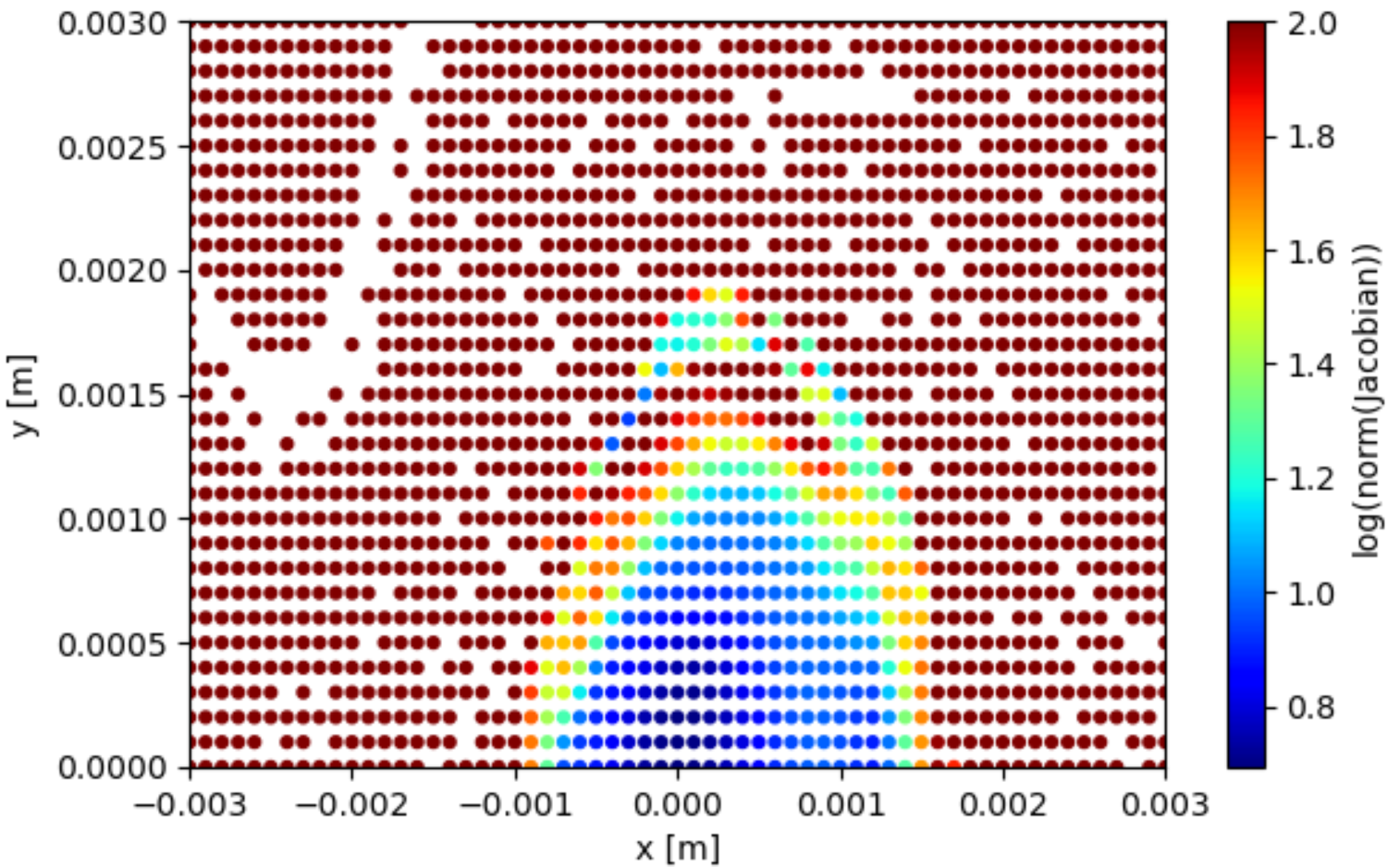}
\includegraphics[width=5.2cm,height=3.1cm]{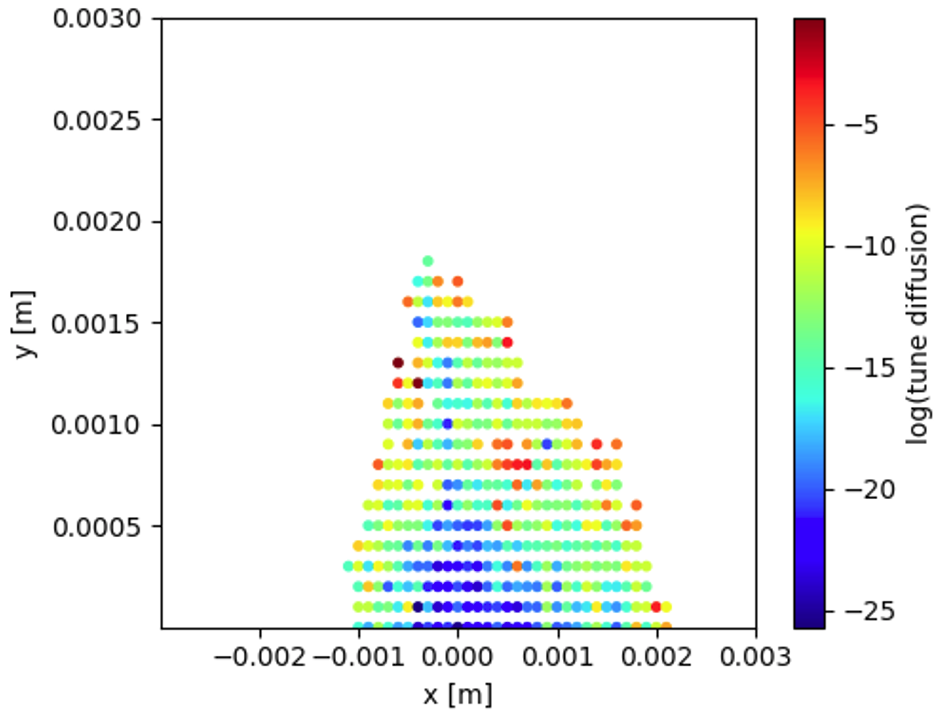}
\caption{
Left: spatial distribution of the log Frobenius norm of the tangent map after one-turn differentiable tracking.  
Right: log tune diffusion rate from FMA, both for the optimized ALS-U lattice with zero momentum deviation.}
\label{fig5}
\end{figure}

Figure~\ref{fig5} presents the spatial distribution of the Frobenius norm of the one-turn tangent map and the tune diffusion rate from FMA for the optimized quadrupole settings. The two indicators show similar phase-space structures. Compared to the spatial distribution (Fig.~\ref{fig3}) from the nominal configuration, which has quadrupole strengths of $(13.80,-13.82)/m^2$, the optimally tuned lattice with settings of $(13.84,-13.77)/m^2$ provides a modest increase in dynamic aperture.
Specifically, the dynamic aperture area determined from the Frobenius norm of the one-turn tangent map via differentiable tracking is 
$1.16\times10^{-6} m^2$
 for the nominal setting, and $1.38\times10^{-6} m^2$ for the optimized setting.

In the above optimization, all sextupoles in the accelerator were kept constant. To explore further improvements, we included two sextupole families located outside the dispersion region alongside the two quadrupole families.
The ranges for these two sextupole families were set to [1, 7] $m^{-3}$ and [-1020, -840] $m^{-3}$, respectively. The resulting optimized dynamic aperture area is $1.42 \times 10^{-6} m^2$, achieved with the optimal settings of $(13.84, -13.77, 4.75, -930.92)$.
Ultimately, including these two sextupole families in the optimization did not significantly improve the dynamic aperture area.

\begin{figure}[!htb]
\centering
\includegraphics[width=5.0cm]{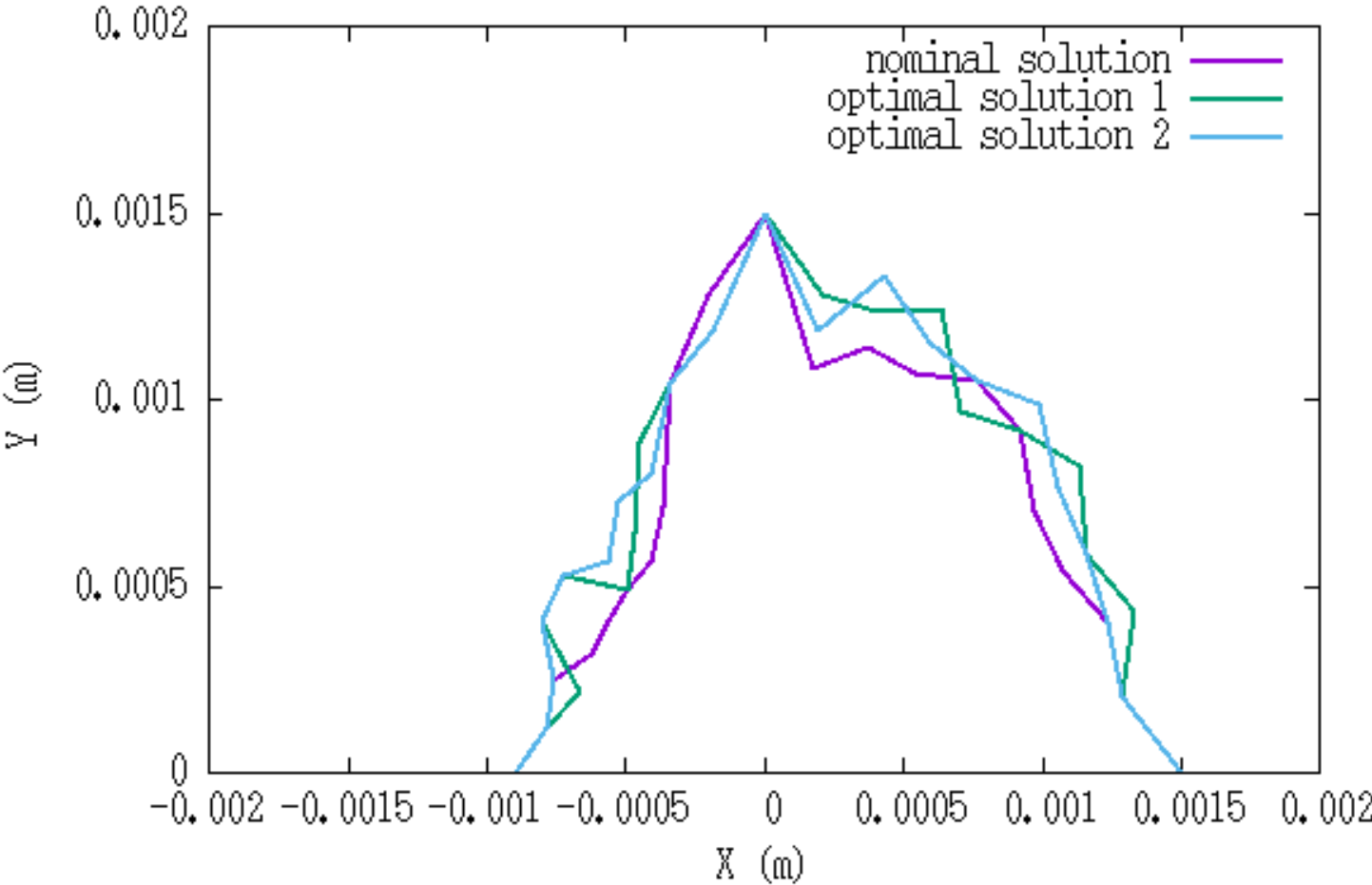}
\caption{
Dynamic aperture boundaries from $1000$-turn particle tracking using the optimized lattice (magenta) and the nominal lattice (green).}
\label{fig6}
\end{figure}

To validate the optimization results, we performed direct particle tracking for $1000$ turns using the optimized quadrupole settings (optimal solution 1) and quadrupole and sextupole settings (optimal solution 2) to obtain the dynamic aperture boundary. Figure~\ref{fig6} compares the boundaries obtained from the nominal and optimized lattices. The optimized configuration yields a noticeably larger dynamic aperture, consistent with the predictions of the tangent map norm indicator.

\section{Conclusions}

In this work, we have proposed the use of the norm of the tangent map, obtained from a short number of turns of particle tracking, as a fast indicator of chaotic trajectories. By employing forward-mode automatic differentiation, the tangent map can be computed automatically within a differentiable particle-tracking framework, without requiring separate derivative calculations.

As an illustrative application, we applied the proposed indicator to dynamic aperture optimization for an ALS-U lattice design. Using the Frobenius norm of the one-turn tangent map as the optimization metric, we obtained an optimal lattice configuration that yields a larger dynamic aperture than the nominal settings.  

These results demonstrate that the tangent map norm indicator, enabled by automatic differentiation, is a practical and efficient tool for the study of nonlinear beam dynamics and for accelerator lattice design optimization.

The norm of the one-turn tangent map obtained from automatic differentiation can serve as an effective indicator for fast initial optimization. However, these initial optimized parameters should be verified or further refined using more accurate, albeit slower, methods such as long-term direct or differentiable tracking, or MFA.

{Furthermore, other effective methods, such as analytical-based resonance correction~\cite{forest,bengtsson} and direct tracking with a flood-fill pattern~\cite{riemann}, are also used for dynamic aperture optimization. The method proposed in this paper can be viewed as an alternative approach to these techniques.}

\section*{ACKNOWLEDGEMENTS}
We would like to thank ALS-U team for one version of the ALS-U design lattice and Dr. M. Venturini for helpful discussion 
about the ALS-U application. This was supported by the U.S. Department of Energy under Contract No. DE-AC02-05CH11231 and 
DE-SC0024170, and used computer resources at the National Energy Research Scientific Computing Center (NERSC).

	{%
	

} 
%
%



\begin{thebibliography}{9} 

\bibitem{borland}M. Borland, L. Emery, V. Sajaev, and A. Xiao, Direct
methods of optimization of storage ring dynamic and
momentum aperture, in Proceedings of the Particle Accelerator
Conference (PAC09), Vancouver, BC, Canada
(JACoW, Geneva, 2009).
\bibitem{sun11}C. Sun, D. Robin, H. Nishimura, C. Steier and W. Wan, Dynamic aperture optimization using genetic algorithms, in Proc. 2011 Particle Accelerator Conference, 
New York, NY, USA, p. 793, 2011.
\bibitem{yang}L. Yang, Y. Li, W. Guo, S. Krinsky, Multiobjective optimization of dynamic
aperture, Phys. Rev. ST Accel. Beams 14 (2011) 054001.
\bibitem{gao}W. Gao, L. Wang, and W. Li, Simultaneous optimization of
beam emittance and dynamic aperture for electron storage
ring using genetic algorithm, Phys. Rev. ST Accel. Beams
14, 094001 (2011).
\bibitem{huang}X. Huang and J. Safranek, Nonlinear dynamics optimization
with particle swarm and genetic algorithms for
SPEAR3 emittance upgrade, Nucl. Instrum. Methods Phys.
Res., Sect. A 757, 48 (2014).

\bibitem{li2018}Y. Li,W. Cheng, L.H. Yu, and R. Rainer, Genetic algorithm
enhanced by machine learning in dynamic aperture optimization,
Phys. Rev. Accel. Beams 21, 054601 (2018).
\bibitem{kran}M. Kranjčević, Multiobjective optimization of the dynamic
aperture using surrogate models based on artificial neural
networks, Phys. Rev. Accel. Beams 24, 014601 (2021).
\bibitem{wan2022}J. Wan and Y. Jiao,
Machine learning enabled fast evaluation of dynamic aperture for storage ring accelerators,
New Journal of Physics 24.6 (2022), p. 063030.
\bibitem{croce}D. Di Croce, M. Giovannozzi, E. Krymova, T. Pieloni, S. Redaelli, M. Seidel,
R. Tomás and F.F. Van der Veken,
Optimizing dynamic aperture studies with active learning, JJournal of
Instrumentation 19, P04004, (2024).


\bibitem{laskar1}J. Laskar and D. Robin, Application of frequency map
analysis to the ALS, in International workshop on single
particle effects in large hadron colliders, LHC’95, Montreux,
Switzerland (1996), 
\bibitem{laskar2}J. Laskar, Introduction to frequency map analysis, in
Hamiltonian Systems with Three or More Degrees of
Freedom, edited by C. Simó, NATO ASI Series (Series
C:Mathematical and Physical Sciences) Vol. 533 (Springer,
Dordrecht, 1999).
\bibitem{steier}C. Steier, W. Wan, Quantitative lattice optimization using frequency map
analysis, Proc. of IPAC (2010) 4746–4748.
\bibitem{sun}C. Sun, D. S. Robin, H. Nishimura, C. Steier and W. Wan,
Small-emittance and low-beta lattice design and optimizations, Phys. Rev. ST Accel. Beams, vol. 15, p. 054001, 2012
\bibitem{yonnis}Y. Papaphilippou, Detecting chaos in particle accelerators through the frequency
map analysis method, Chaos 24 (2) (2014) 024412.
\bibitem{xu}H. S. Xu, W. H. Huang, C. X. Tang, and S. Y. Lee,
Design of a 4.8-m ring for inverse Compton scattering
x-ray source, Phys. Rev. Accel. Beams 17, 070101
(2014).
\bibitem{hwang}K. Hwang, C. Mitchell, R. Ryne, Rapidly converging chaos indicator for studying
dynamic aperture in a storage ring with space charge, Phys. Rev. Accel. Beams
23 (2020) 084601
\bibitem{li2}Y. Li, Y. Hao, K. Hwang, R. Rainer, A. He, A. Liu,
Fast dynamic aperture optimization with forward-reversal integration, 
Nucl. Instrum. and Methods A 988 (2021) 164936.

\bibitem{ad}
C. C. Margossian, \textquotedblleft{A Review of Automatic Differentiation and its Efficient Implementation.}\textquotedblright
Wiley interdisciplinary reviews: data mining and knowledge discovery 9, no. 4 (2019): e1305. 
\url{https://doi.org/10.1002/widm.1305}
\bibitem{pytorch}
A. Paszke et al., 
\textquotedblleft{PyTorch: An Imperative Style,
High-Performance Deep Learning Library,}\textquotedblright in Advances
in Neural Information Processing Systems 32 , edited by
H. Wallach, H. Larochelle, A. Beygelzimer, F. d. Alche-Buc, 
E. Fox, and R. Garnett (Curran Associates, Inc.,2019) pp. 8024-8035.
\bibitem{tensorflow}
Martin Abadi et al., 
\textquotedblleft{TensorFlow: Large-scale machine learning on heterogeneous systems,}\textquotedblright
2015.
\url{https://www.tensorflow.org/}   

\bibitem{roussel2022}
R. Roussel, A. Edelen, D. Ratner, K. Dubey, J. P. Gonzalez-Aguilera, Y.K. Kim, and N. Kuklev, 
Differentiable Preisach modeling for characterization and optimization of particle accelerator systems with hysteresis,
\textquotedblright Phys. Rev. Lett. 128, 204801 (2022).
\bibitem{roussel2023b}R. Roussel, A. Edelen, C. Mayes, D. Ratner, J. P. Gonzalez-Aguilera, S. Kim, E. Wisniewski, and J. Power, 
Phase space reconstruction from accelerator beam measurements using neural networks and differentiable simulations,
Phys. Rev. Lett. 130, 145001 (2023).
\bibitem{qiang2023}J. Qiang, 
Differentiable self-consistent space-charge simulation for accelerator design,
\textit{Phys. Rev. Accel Beams}, vol. 26, 024601, 2023. 
\bibitem{cheetah}J. Kaiser, C. Xu, A. Eichler, and A. S. Garcia,
Bridging the gap between machine learning and particle accelerator
physics with high-speed, differentiable simulations,
\textit{Phys. Rev. Accel Beams}, vol. 27, 054601, 2024. 
\bibitem{wan2}J. Wan, H. Alamprese, C. Ratcliff, J. Qiang, Y. Hao, JuTrack: a Julia package for auto-differentiable accelerator modeling and particle tracking, 
\textit{Comp. Phys. Comm.} 309, 109497 (2025).

\bibitem{berz}M. Berz, Differential Algebraic Description of Beam Dynamics to Very High Orders, Part. Accel. 24, 109 (1989).

\bibitem{qiangipac25}J. Qiang, Y. Hao, A. Qiang, J. Wan, A module for fast auto differentiable simulations, in Proc. of IPAC25, WEBN2, p.1671, 2025. (\url{https://github.com/qianglbl/mAD})

\bibitem{adwiki}\url{https://en.wikipedia.org/wiki/Automatic\_differentiation}    
\bibitem{liapounoff}A. M. Lyapunov, Problème général de la stabilité du mouvement, in: Annales de la
Faculté Des Sciences de Toulouse: Mathématiques, Vol. 9, 1907, pp. 203–474.

\bibitem{henon}M. Henon and C. Heiles, The applicability of the third
integral of motion: Some numerical experiments, Astron. J.
69, 73 (1964).

\bibitem{dragt}A. J. Dragt, and J. M. Finn, Lie Series and Invariant Functions for Analytic Symplectic Maps. Journal of Mathematical Physics, 17, 2215 (1976).
\bibitem{alsu}C. Steier, et al., Status of the Conceptual Design of ALSU,
in Proc. of IPAC’17, Copenhagen, Denmark, May 2017,
p. 2824, 2017.
\bibitem{elegant}M. Borland, Elegant: A Flexible SDDS-Compliant Code for Accelerator Simulation,
Tech. Rep. LS-287, Argonne National Lab., IL, US, 2000.

\bibitem{qiangpvp}
J. Qiang, A parallel variable population multi-objective optimizer for accelerator beam dynamics optimization, Nuclear Inst. and Methods in Physics Research, A 1054 (2023) 168402.


\bibitem{qiangde}J. Qiang, C. Mitchell, A. Qiang, A Tuning of an adaptive unified differential evolution
algorithm for global optimization, in: Proceedings of CEC2016, Vancouver,
Canada, 24-29 2016, pp. 4061–4068.

\bibitem{deb}K. Deb, Multi-Objective Optimization using Evolutionary Algorithms, John Wiley
and Sons, Chichester, UK, 2001.

\bibitem{forest}E. Forest, M. Berz and J. Irwin, Normal form methods for complicated periodic systems : a complete solution using differential algebra and lie operators, Part. Accel., 24, 91-113 (1989).
\bibitem{bengtsson}J. Bengtsson, The Sextupole Scheme for the Swiss Light Source (SLS): An Analytic
Approach, SLS Technical Report No. 9/97, 1997

\bibitem{riemann}B. Riemann, M. Aiba, J. Kallestrup, and A. Streun, Efficient algorithms for dynamic aperture and momentum acceptance calculation in synchrotron light sources, Phys. Rev. Accel. Beams 27, 094002 (2024).

	\end{thebibliography}
\end{document}